\documentclass[12pt]{amsart}
\usepackage[dvips]{graphicx}
\newtheorem{thm}{Theorem}[section]

\newtheorem{prop}[thm]{Proposition}
\theoremstyle{definition}

\theoremstyle{remark}
\newtheorem{rem}[thm]{Remark}
\numberwithin{equation}{section}

\newcommand{\Real}{\mathbb R}

\newcommand{\To}{\longrightarrow}

\newcommand{\C}{\mathbb{C}}
\newcommand{\M}{\mathbb{M}}

\newcommand {\sqr}{\hspace{\fill}$\Box$}

\begin{document}

\title{An Alternative Mathematical Model For Special Relativity}

\author{Guy Tsabary and Aviv Censor}
\address{Department of Mathematics, Technion, Haifa 32000, Israel}

\email{tsabary@tx.technion.ac.il, avivc@tx.technion.ac.il}
\date{\today}

\begin{abstract}

We present a mathematical model for a physical theory that is
compatible with Einstein's Special Relativity Theory. Our model
consists of three pseudo-complex dimensions, representing three
real dimensions of space, dual to what could be interpreted as
three real dimensions of time. We use the term "pseudo-complex",
since the mathematical object representing each paired space-time
coordinate in our model, is a two-dimensional space over
$\mathbb{R}$ whose multiplication makes it into a non-commutative
non-associative algebra. We have been unable to find records of
this very elementary mathematical object in the literature.

\end{abstract}
\maketitle

\section{\bf{Introduction}}

In Einstein's Special Relativity Theory, a physical event may be
represented by a 4-dimensional vector of the form
\[ \vec{\chi} = \left( \begin{array}{c} x \\ y \\ z \\ cti
\end{array} \right) . \]
Passing to another frame of reference is done using the Lorentz
transformation, which is given by a $4 \times 4$ matrix $\Lambda$
(see below), by $\vec{\chi} \ ' = \Lambda \vec{\chi}$. In our
proposed alternative model, a physical event is represented by a
vector of the form
\[ \vec{x} = \left( \begin{array}{c}
xe + t_xi \\ ye + t_yi \\ ze + t_zi
\end{array} \right) . \]
While $x,y,$ and $z$ are the regular space coordinates, $t_x,t_y,$
and $t_z$ are derived from the single time coordinate $t$, and
satisfy $(t_x^2+t_y^2+t_z^2) = t^2$. Each pair of space and time
coordinates are then coupled into one "pseudo-complex" coordinate.
More precisely, $e$ and $i$ form a basis for a two dimensional
vector space over $\Real$, which we denote by $\M$, whose elements
are thus of the form $\alpha e + \beta i$. We call $\M$
"pseudo-complex" since we endow it with a multiplication differing
from the one in $\mathbb{C}$, by defining
\[ ee = e, ~~ei = i, ~~ie = -i, ~~ii = -e. \]
$\M$ turns out to be non-commutative and non-associative. A brief
account of the underlying mathematics of $\M$ appears in Section
\ref{math}.

Section \ref{phys} introduces Special Relativity in terms of our
proposed model. The Lorentz transformation in our model is given
by a matrix $L$ in $\M_{3 \times 3}$ (see below), constructed so
that the transformation $\vec{x} \ ' = L \cdot \vec{x}$ is
consistent with the regular Lorentz transformation above. \\ We
then demonstrate other known physical entities such as the
current-charge density vector, the electromagnetic field tensor,
and the angular momentum tensor.

In section \ref{results} we present physical results obtained
using our model. We first introduce two assumptions, that are
motivated by the $\M$ mathematics. The assumptions are that a
vector field $\vec{f}$ in $\M^3$, given by
\[ \vec{f} = \left( \begin{array}{c} f_xe + g_xi \\ f_ye + g_yi
\\ f_ze + g_zi \end{array} \right) , \]
satisfies
\[ \sum_{i=1}^{3} \left( \frac{\partial f_i}{\partial x_i} +
\frac{\partial g_i}{\partial t_{x_i}} \right) = 0 \hspace{1cm}
\text{and} \hspace{1cm} \sum_{i=1}^{3} \left( \frac{\partial
f_i}{\partial t_{x_i}} +
\frac{\partial g_i}{\partial x_i} \right) = 0. \] %
Based on these assumptions, we obtain in Theorem \ref{Maxwell} the
continuity equation $$\nabla \cdot \vec{j}= -\frac{\partial
\rho}{\partial t}$$ and the curl Maxwell equations $$\nabla \times
\vec{E} = - \frac{\partial \vec{B}}{\partial t} \hspace{1cm}
\text{and} \hspace{1cm} \nabla \times \vec{B} =
\frac{\partial \vec{E}}{\partial t}.$$ %
In Theorem \ref{radial} we prove, based on the same assumptions,
that the velocity $s$ of a particle in a radial force field $F$
satisfies a certain partial differential equation, namely:
\[ \gamma_s ~ \frac{\partial{s}}{\partial{t}} + s ~
\frac{\partial{\gamma_s}}{\partial{t}} + \frac{\partial
\gamma_s}{\partial r} = 0. \]

Finally, in Section \ref{remarks}, we discuss several aspects of
our theory which deserve further investigation. These include more
$\M$-motivated assumptions at the rank-2 tensor level, that lead
to all four Maxwell equations.

\section{\bf{The underlying mathematics}}\label{math}

Let $\M$ be a two dimensional vector space over $\Real$. Let
$\{e,i\}$ be a basis for $\M$. We introduce a multiplication on
$\M$ using the bilinear map $\M \times \M \To \M$ defined by:
\[ ee = e, ~~ei = i, ~~ie = -i, ~~ii = -e. \]
It is clear that $\M$ is non-commutative, since $ei \neq ie$.
Simple calculations show that $\M$ is also non-associative and
non-unital. Note that the difference between $\M$ and $\C$ can be
seen in the definition $ie=-i$ in $\M$, whereas in $\C$, $i1=i$.
Since $\{e,i\}$ is a basis for $\M$, every element $x \in \M $ has
a unique representation $x=\alpha e + \beta i, \ \alpha, \beta \in
\Real.$ We will denote $Re(x):=\alpha, \ Im(x):= \beta,$ and thus
$x=Re(x)e+Im(x)i.$ Notice that the bilinearity of the
multiplication translates to the distributivity laws:
\[(x+y)z=xz+yz \ \ and \ \ x(y+z)=xy+xz \ \ for \ all \ x,y,z \in
\M.\] We also define a map $*:\M \To \M$ by $x^*:=xe$. This will
play the role of the complex conjugate.

The properties in the following proposition are not difficult to
verify:
\begin{prop}\label{properties}
For all $x,y,z$ in $\M$:
\[
\begin{array}{ll}
1. & \alpha Re(x)=Re(\alpha x),~~\alpha Im(x)=Im(\alpha x),~\forall~\alpha \in \Real. \\ %
2. & \text{e is a left unit element, that is}: ex=x,~\forall~ x \in \M. \\ %
3. & Re(e)\!\!=\!1,~Im(e)\!\!=\!0,~Re(i)\!\!=\!0,~Im(i)\!\!=\!1,~Re(0)\!\!=\!0,~Im(0)\!\!=\!0. \\ %
4. & e^*=e, \\ %
   & i^*=-i. \\ %
5. & x^*=Re(x)e-Im(x)i. \\ %
6. & Re(x^*)=Re(x), \\ %
   & Im(x^*)=-Im(x). \\ %
7. & Re(xy)=Re(x)Re(y)-Im(x)Im(y), \\ %
   & Im(xy)=Re(x)Im(y)-Im(x)Re(y). \\ %
8. & x(yz)=y(xz). \\ %
9. & x^*(yz)=(xy)z. \\ %
10. & (xy)z=(zy)x. \\ %
11. & x-x^*=2Im(x)i, \\ %
    & x+x^*=2Re(x)e. \\ %
12. & x^{**}=x. \\ %
13. & xy^*=yx^*,~and~x^*y=y^*x. \\ %
14. & xy=y^*x^*=(yx)^*. \\ %
15. & Im(x^2)=0~and~Re(x^2)=Re(x)^2-Im(x)^2.
\end{array}
\]
\end{prop}

Elements $x \in \M$ with the property that $Re(x) \neq
\pm{Im(x)}$, can be seen to fulfill the condition $Re(x^2) \neq
0$. These play a role of invertible elements, since if we define
$x^{-1}:= \frac{1}{Re(x^2)}x$, then $x^{-1}$ satisfies the
property $x^{-1}x = xx^{-1} = e$ (although $e$ is merely a left
unit).

We define matrix multiplication $A \cdot B: \M_{n \times n} \times
\M_{n \times n} \longrightarrow \M_{n \times n}$ by $$(A \cdot
B)_{i,j}:= (A^*B)_{i,j} = \sum_{k=1}^{n}a_{ik}^* b_{kj}.$$
Likewise, the definition of an $n \times n$ matrix acting on a
vector $\vec{x} \in \M^n$ is: $$(A \cdot \vec{x})_i:=
(A^*\vec{x})_{i} = \sum_{k=1}^{n}a_{ik}^* x_k.$$ Under these
definitions, matrices are linear operators (over $\mathbb{R}$),
and matrix multiplication is consistent with composition of
operators, and is therefore associative.

\section{\bf{Special relativity in terms of the proposed model}}\label{phys}

We now turn to the Physics. In Einstein's Theory, a physical event
is represented by a 4-dimensional vector of the form
\[ \vec{\chi} = \left( \begin{array}{c} x \\ y \\ z \\ cti
\end{array} \right) . \]
Denote the same event, when viewed in a second frame of reference,
by
\[ \vec{\chi} \ ' = \left(\begin{array}{c} x' \\ y' \\ z' \\ ct'i
\end{array} \right) . \]
Suppose the second frame of reference is moving with velocity
$\vec{v}=(v_x,v_y,v_z)$ relative to the first. Then the well known
Lorentz transformation $\Lambda$ relates the two, according to
\[ \vec{\chi} \ ' = \Lambda \vec{\chi} . \]
$\Lambda$ is given explicitly by
\[ \Lambda = \left( \begin{array}{cccc} %
1 + (\gamma-1)\alpha_x^2 & \alpha_x\alpha_y(\gamma - 1) &
\alpha_x\alpha_z(\gamma - 1) & \beta\gamma\alpha_x i \\ %
\alpha_x\alpha_y(\gamma - 1) & 1 + (\gamma-1)\alpha_y^2 &
\alpha_y\alpha_z(\gamma - 1) & \beta\gamma\alpha_y i \\ %
\alpha_x\alpha_z(\gamma - 1) & \alpha_y\alpha_z(\gamma - 1) &
1 + (\gamma-1)\alpha_z^2 & \beta\gamma\alpha_z i \\ %
-\beta\gamma\alpha_x i & -\beta\gamma\alpha_y i &
-\beta\gamma\alpha_z i & \gamma %
\end{array}\right) , \]
where $\alpha_x:=\frac{v_x}{v}, \ \alpha_y:=\frac{v_y}{v}, \
\alpha_z:=\frac{v_z}{v},$ and $v:=|\vec{v}|=(v_x^{2} + v_y^{2} +
v_z^{2})^{\frac{1}{2}}.$ As usual, $\beta:= \frac{v}{c}$ and
$\gamma:= \frac{1}{\sqrt{1-\beta^{2}}}.$

In our model, a physical event is represented by a three
dimensional vector in $\M^{3}$, of the form
\[ \vec{x} = \left( \begin{array}{c}
xe + t_xi \\ ye + t_yi \\ ze + t_zi
\end{array} \right) , \]
where $t_x:=\alpha_x c t, \  t_y:=\alpha_y c t,$ and
$t_z:=\alpha_z c t$. Thus, a physical event in a certain frame of
reference $A$, is not uniquely represented. On the contrary: it's
representation depends on a second frame of reference $B$, moving
with velocity $\vec{v}_{A,B} \neq 0$ relative to $A$. An event in
$A$ is represented differently with respect to each frame $B$.

We replace the usual Lorentz transformation $\Lambda$ with the
following matrix $L \in \M_{3 \times 3}$:

\[ L = \left( \begin{array}{rrr}  \, %
[1 + (\gamma-1)\alpha_x^2]e & [\alpha_x\alpha_y(\gamma - 1)]e &
[\alpha_x\alpha_z(\gamma - 1)]e \\  \, %
-[\beta\gamma\alpha_x^2]i & -[\alpha_x\alpha_y\beta\gamma]i &
-[\alpha_x\alpha_z\beta\gamma]i \vspace{0.2cm} \\ \,
[\alpha_x\alpha_y(\gamma - 1)]e  & [1 + (\gamma-1)\alpha_y^2]e &
[\alpha_y\alpha_z(\gamma - 1)]e \\ \, %
-[\alpha_x\alpha_y\beta\gamma]i & -[\beta\gamma\alpha_y^2]i &
-[\alpha_y\alpha_z\beta\gamma]i \vspace{0.2cm} \\ \,
[\alpha_x\alpha_z(\gamma - 1)]e & [\alpha_y\alpha_z(\gamma - 1)]e
& [1 + (\gamma-1)\alpha_z^2]e \\ \, %
-[\alpha_x\alpha_z\beta\gamma]i & -[\alpha_y\alpha_z\beta\gamma]i
& -[\beta\gamma\alpha_z^2]i
\end{array} \right) .\]
We define $$\vec{x} \ ' := L \cdot \vec{x}.$$ By calculating the
components explicitly and comparing, one can verify that $\vec{x}
\ '$ is consistent with the usual Lorentz transformation result
$\vec{\chi} \ '$, that is:
\[ \vec{x} \ ' = \left(
\begin{array}{c} x'e + t_x' i \\ y'e + t_y' i \\ z'e + t_z' i
\end{array} \right) , \] %
where $t_x'=\alpha_x c t', \ t_y'=\alpha_y c t',$ and
$t_z'=\alpha_z c t'.$ Note though that $\alpha_x, \alpha_y,
\alpha_z$ in $\vec{x} \ '$ fit the original velocity
$\vec{v}_{A,B}$. That is, if the regarded physical event would
have been represented in the frame $B$ with respect to the frame
$A$ (which is moving with velocity $\vec{v}_{B,A}=-\vec{v}_{A,B}$
relative to $B$), then the representation would be $(\vec{x} \
')^*$ rather than $\vec{x} \ '$.

We further demonstrate our model using examples of other physical
entities such as the current-charge density vector, the
electromagnetic field tensor, and the angular momentum tensor.

The current-charge density vector is classically represented by
\[ \vec{d} = \left( \begin{array}{c}
j_x \\ j_y \\ j_z \\ \rho i
\end{array} \right) .\] %
$\vec{d} \, '$ is then obtained by $\vec{d} \, ' = \Lambda
\vec{d}$. \\ In our model the current-charge density vector is
represented by \[ \vec{\delta} = \left( \begin{array}{c} j_xe +
\alpha_x \rho i \\ j_ye + \alpha_y \rho i \\ j_ze + \alpha_z \rho
i \end{array} \right) ,\] %
and $\vec{\delta} \, '  = L \cdot \vec{\delta}.$ One can verify
that $\vec{\delta} \, '$ is consistent with $\vec{d} \, '$ in the
same sense as before.

The electromagnetic field is classically represented by a tensor
of rank 2:
\[ F = \left( \begin{array}{cccc} %
0 & B_z & -B_y & -E_x i \\ %
-B_z & 0 & B_x & -E_y i \\ %
B_y & -B_x & 0 & -E_z i \\ %
E_x i & E_y i & E_z i & 0 %
\end{array}\right) ,\]
where $B_x, B_y, B_z$ and $E_x, E_y, E_z$ are the magnetic and
electric components, respectively. $F'$ is then obtained by
$\Lambda F \Lambda^{-1}.$ \\ In our model the electromagnetic
field can be represented by a vector in $\M^3$ (a rank 1 tensor),
of the following form:

\[ \vec{\mathcal{F}} \!=\!\! \left( \!\!\! \begin{array}{l} \, %
[\alpha_x(B_y + B_z) + \alpha_yB_y + \alpha_zB_z]e + \\ \, %
[\alpha_x^2(E_y \!\!-\!\! E_z) \!+\! \alpha_y^2E_y \!-\!
\alpha_z^2E_z \!-\! \alpha_x\alpha_yE_z \!+\! \alpha_x\alpha_zE_y
\!+\! \alpha_y\alpha_z(E_z \!\!-\!\! E_y)]i \vspace{0.2cm} \\ \,
[-\alpha_xB_x + \alpha_y(B_z-B_x) - \alpha_zB_z]e + \\ \, %
[-\alpha_x^2E_x \!-\! \alpha_y^2(\! E_x \!\!+\!\! E_z \! )  \!-\!
\alpha_z^2E_z \!-\! \alpha_x\alpha_yE_z \!-\!
\alpha_x\alpha_z(\!E_x \!\!+\!\! E_z\!) \!+\!
\alpha_y\alpha_zE_x]i \vspace{0.2cm} \\ \,
[-\alpha_xB_x - \alpha_yB_y - \alpha_z(B_x-B_y)]e + \\ \, %
[\alpha_x^2E_x \!+\! \alpha_y^2E_y \!+\! \alpha_z^2(E_x \!\!+\!\!
E_y) \!+\! \alpha_x\alpha_y(E_x \!\!+\!\! E_y) \!+\!
\alpha_x\alpha_zE_y \!-\! \alpha_y\alpha_zE_x]i
\end{array} \!\!\!\! \right) \!\! . \] %
Since $\vec{\mathcal{F}}$ is a vector, $\vec{\mathcal{F}} \, '$ is
obtained by $\vec{\mathcal{F}} \, ' = L \cdot \vec{\mathcal{F}}.$
Once again, $\vec{\mathcal{F}} \, '$ is consistent with $F \, '$,
in the sense that if we denote \[ F' = \left( \begin{array}{cccc} %
0 & B'_z & -B'_y & -E'_x i \\ %
-B'_z & 0 & B'_x & -E'_y i \\ %
B'_y & -B'_x & 0 & -E'_z i \\ %
E'_x i & B'_y i & B'_z i & 0 %
\end{array}\right), \] %
then a very tedious calculation proves that $\vec{\mathcal{F}} \,
'$ is exactly
\[ \vec{\mathcal{F}} \, \! ' \!\!=\!\! \left( \!\!\!\! %
\begin{array}{l} \,
[\alpha_x(B_y' + B_z')  + \alpha_yB_y' + \alpha_zB_z']e + \\ \, %
[\alpha_x^2 \!(\! E_y' \!-\! E_z' \!) \!+\! \alpha_y^2E_y' \!-\!
\alpha_z^2E_z' \!-\! \alpha_x\alpha_yE_z' \!+\!
\alpha_x\alpha_zE_y' \!+\! \alpha_y\alpha_z \!(\! E_z' \!-\! E_y'
\!) ]i \vspace{0.1cm} \\ \,
[-\alpha_xB_x' + \alpha_y(B_z'-B_x') - \alpha_zB_z']e + \\ \, %
[-\alpha_x^2E_x'  \!-\! \alpha_y^2 \!(\! E_x' \!+\! E_z' \!) \!-\!
\alpha_z^2E_z' \!-\! \alpha_x\alpha_yE_z' \!-\! \alpha_x\alpha_z
\!(\! E_x' \!+\! E_z' \!) \!+\! \alpha_y\alpha_zE_x']i
\vspace{0.1cm} \\ \,
[-\alpha_xB_x' - \alpha_yB_y' - \alpha_z(B_x'-B_y')]e + \\ \, %
[\alpha_x^2E_x' \!+\! \alpha_y^2E_y' \!+\! \alpha_z^2 \!(\! E_x'
\!+\! E_y' \!) \!+\! \alpha_x\alpha_y \!(\! E_x' \!+\! E_y' \!)
\!+\! \alpha_x\alpha_zE_y' \!-\! \alpha_y\alpha_zE_x']i
\end{array} \!\!\!\! \right) \!\! . \] %
The electromagnetic field in our model also has a representation
as a tensor of rank 2, namely:
\[ \mathcal{F} = \left( \begin{array}{ccc} %
0 & B_z e + & -B_y e + \\ \, %
&[\alpha_x E_y - \alpha_y E_x]i & [\alpha_x E_z - \alpha_z E_x]i
\vspace{0.1cm} \\
-B_z e + & 0 & B_x e + \\ \, %
[\alpha_y E_x - \alpha_x E_y]i & & [\alpha_y E_z - \alpha_z E_y]i
\vspace{0.1cm} \\
B_y e + & -B_x e + &  0 \\ \, %
[\alpha_z E_x - \alpha_x E_z]i & [\alpha_z E_y - \alpha_y E_z]i &
\end{array}\right)  \]
As such, $\mathcal{F} '$ is obtained by $\mathcal{F} ' = L \cdot
\mathcal{F} \cdot L$ (see Remark \ref{Fmunu} below). Consistency
with $F'$ can be verified here too.

Angular momentum is classically represented by the following
tensor of rank 2:
\[ J = \left( \begin{array}{cccc} %
0 & x P_y - y P_x & x P_z - z P_x & (x E - P_x t) i \\ %
y P_x - x P_y & 0 & y P_z - z P_y & (y E - P_y t) i \\ %
z P_x - x P_z & z P_y - y P_z & 0 & (z E - P_z t) i \\ %
-(x E - P_x t) i & -(y E - P_y t) i & (z E - P_z t) i & 0 %
\end{array}\right) , \] %
where $P_x, P_y, P_z$ are the components of the angular momentum,
and $E$ is the energy. $J'$ is obtained by $J'= \Lambda J
\Lambda^{-1}.$ \\ %
In our model, angular momentum has a representation as a tensor of
rank 2, namely:
\[ \mathcal{J} = \left( \begin{array}{ccc} %
& (x P_y - y P_x)e + & (x P_z - z P_x)e + \\ \, %
0 &[\alpha_y(x E - P_x t) -  & [\alpha_z(x E - P_x t) -  \\ \, %
& \alpha_x(y E - P_y t)]i & \alpha_x(z E - P_z t)]i \vspace{0.1cm} \\
(y P_x - x P_y)e + & & (y P_z - z P_y)e + \\ \, %
[\alpha_x(y E - P_y t) -  & 0 & [\alpha_z(y E - P_y t) - \\ \, %
\alpha_y(x E - P_x t)]i & & \alpha_y(z E - P_z t)]i \vspace{0.1cm} \\
(z P_x - x P_z)e + & (z P_y - y P_z)e + &  \\ \, %
[\alpha_x(z E - P_z t) - & [\alpha_y(z E - P_z t) - & 0 \\ \,
\alpha_z(x E - P_x t)]i & \alpha_z(y E - P_y t)]i &
\end{array}\right) . \]
Similar to $\mathcal{F} ' $, $\mathcal{J} ' $ is obtained by
$\mathcal{J} ' = L \cdot \mathcal{J} \cdot L$, and consistency
with $J'$ can be verified. Angular momentum also has a vector
representation.

\section{\bf{Obtaining physical results using the proposed model}}\label{results}

A vector field $\vec{f}$ in $\M^3$ is given by %
\[ \vec{f} = \left( \begin{array}{c} f_1 \\ f_2 \\ f_3
\end{array} \right) , \] %
where each $f_i$ is a function $f_i: \M^3 \longrightarrow \M.$ We
can also write $\vec{f}$ as
\[ \vec{f} = \left( \begin{array}{c} f_xe + g_xi \\ f_ye + g_yi
\\ f_ze + g_zi \end{array} \right) , \] %
where now each of $f_x, g_x, f_y, g_y, f_z, g_z$ are thought of as
real functions of the variables $x, t_x, y, t_y, z,$ and $t_z$.

Let $\vec{f}$ be such a vector field. We require the following two
assumptions, by which we restrict ourselves to vector fields in
$\M^3$ originating from vector fields in $\mathbb{R}^4$
representing known physical entities:

\begin{enumerate}
\item
$\vec{f}$ is a function of $t:=\sqrt{t_x^2 + t_y^2 + t_z^2} \ \ .$
\item
$\vec{f}$ satisfies $L \cdot \vec{f} = \vec{f}',$ such that
$\vec{f}'$ is consistent with the usual Lorentz transformation, as
in the examples above.
\end{enumerate}
We now add two more assumptions on the vector field $\vec{f}$:
\[ (a1) \hspace{1cm} \sum_{i=1}^{3} \left( \frac{\partial f_i}{\partial x_i} + %
\frac{\partial g_i}{\partial t_{x_i}} \right) = 0. \]
\[ (a2) \hspace{1cm} \sum_{i=1}^{3} \left( \frac{\partial f_i}{\partial t_{x_i}} + %
\frac{\partial g_i}{\partial x_i} \right) = 0. \] %
One can verify that both the above sums are invariants of $L$,
i.e.
\[ \sum_{i=1}^{3} \left( \frac{\partial f_i '}{\partial x_i '} + %
\frac{\partial g_i'}{\partial t_{x_i}'} \right) = %
\sum_{i=1}^{3} \left( \frac{\partial f_i}{\partial x_i} + %
\frac{\partial g_i}{\partial t_{x_i}} \right) , \] %
and
\[ \sum_{i=1}^{3} \left( \frac{\partial f_i'}{\partial t_{x_i}'} + %
\frac{\partial g_i'}{\partial x_i'} \right) = %
\sum_{i=1}^{3} \left( \frac{\partial f_i}{\partial t_{x_i}} + %
\frac{\partial g_i}{\partial x_i} \right) . \] %
The assumptions $(a1)$ and $(a2)$, that all these sums are equal
to zero, are motivated by the $\M$ mathematics: %
A function $f:\M \To \M$ can be written as $f(xe + yi) = u(x,y)e +
v(x,y)i$ where $u,v:\Real^2 \To \Real$. If $f$ satisfies a certain
differentiability condition, then u and v both have partial
derivatives, and the following versions of the Cauchy-Riemann
equations hold: $\frac{\partial u}{\partial x} = - \frac{\partial
v}{\partial y}$ and $\frac{\partial u}{\partial y} = -
\frac{\partial v}{\partial x}$.

Our first results deal with Electrodynamics. Clearly, our
current-charge density vector and electromagnetic field vector
satisfy assumptions (1) and (2). We claim that in inertial
systems, by imposing the assumptions $(a1)$ and $(a2)$, we obtain
the continuity equation for current-charge density, and the curl
Maxwell equations for the electromagnetic field (in vacuum,
without sources). We state and prove this claim formally as a
Theorem.

\begin{thm}\label{Maxwell}
1. Suppose that the current-charge density vector $\vec{\delta}$
satisfies assumption $(a1)$. Then
\begin{equation}\label{continuity}
\nabla \cdot \vec{j}= -\frac{\partial \rho}{\partial t}.
\end{equation}
2. Suppose that the electromagnetic field vector
$\vec{\mathcal{F}}$ satisfies assumptions $(a1)$ and $(a2)$. Then

\begin{equation}\label{MaxwellE}
\nabla \times \vec{E} = - \frac{\partial \vec{B}}{\partial t}.
\end{equation}
\begin{equation}\label{MaxwellB}
\nabla \times \vec{B} = \frac{\partial \vec{E}}{\partial t}.
\end{equation}
\end{thm}

\textbf{Proof:} 1. Recall that the current-charge density vector
is
\[ \vec{\delta} = \left( \begin{array}{c} j_xe + \alpha_x \rho i \\ %
j_ye + \alpha_y \rho i \\ j_ze + \alpha_z \rho i
\end{array} \right) .\] %
We assume that $\vec{\delta}$ satisfies $(a1)$, so that
\begin{equation}\label{a1_for_cont}
- \sum_{i=1}^{3}\frac{\partial j_{x_i}}{\partial x_i} =  %
\sum_{i=1}^{3} \frac{\partial (\alpha_i \rho) }{\partial t_{x_i}}.
\end{equation}
The left hand side of the above equation is precisely $ - \nabla
\cdot \vec{j}$. We need to calculate the right hand side. Since
$\alpha_i \rho$ is a function of
$t:=(\sum_{i=1}^{3}t_{x_i}^2)^{\frac{1}{2}},$ %
\[ \frac{\partial (\alpha_i \rho) }{\partial t_{x_i}} = %
\frac{\partial t}{\partial t_{x_i}}\frac{\partial (\alpha_i \rho)
}{\partial t} . \] Now, since
\begin{equation}\label{dtdtxi}
\frac{\partial t}{\partial t_{x_i}} =
\frac{\partial((\sum_{i=1}^{3}t_{x_i}^2)^{\frac{1}{2}})}{\partial
t_{x_i}} = \frac{1}{2}
\frac{1}{(\sum_{i=1}^{3}t_{x_i}^2)^{\frac{1}{2}}} 2 t_{x_i} =
\frac{t_{x_i}}{t},
\end{equation}
it follows that
\[ \frac{\partial (\alpha_i \rho) }{\partial t_{x_i}} = %
\frac{t_{x_i}}{t}\frac{\partial (\alpha_i \rho) }{\partial t} . \]
Since we are dealing with inertial systems, $\frac{\partial
\alpha_i}{\partial t} = 0.$ Thus,
\[ \frac{\partial (\alpha_i \rho) }{\partial t_{x_i}} = %
\frac{t_{x_i}}{t} \alpha_i \frac{\partial \rho }{\partial t} .
\]
If we restrict to points in $\M^3$ where
$\frac{t_{x_i}}{t}=\alpha_i$ (for any $t$ there exist such
$t_{x_i}$'s), we obtain
\[ \frac{\partial (\alpha_i \rho) }{\partial t_{x_i}} = %
\alpha_i^2 \frac{\partial \rho }{\partial t}. \] By summing over
$i$, we find that the right hand side of (\ref{a1_for_cont}) is
\[ \sum_{i=1}^{3} \frac{\partial (\alpha_i \rho) }{\partial
t_{x_i}} = \sum_{i=1}^{3} \alpha_i^2 \frac{\partial \rho
}{\partial t} = \frac{\partial \rho }{\partial t} .\] Thus we
obtained the continuity equation (\ref{continuity}). Note that the
result holds for any $x,y,z,t$, and is not affected by the
restriction on the $t_{x_i}$'s.

2. We first impose the assumption $(a2)$ on our electromagnetic
field vector $\vec{\mathcal{F}}$:
\begin{equation}\label{a2_for_elec}
\sum_{i=1}^{3} \left[ \frac{\partial Re(\mathcal{F}_i)}{\partial t_{x_i}} -  %
\frac{\partial Im(\mathcal{F}_i)}{\partial
x_i} \right] = 0 . \end{equation} %
From (\ref{dtdtxi}) we know that $\frac{\partial t}{\partial
t_{x_k}} = \frac{t_{x_k}}{t}$. Adding the fact that
$\frac{\partial \alpha_i}{\partial t} = 0,$ and restricting to
points where $\frac{t_{x_k}}{t}=\alpha_k$, we find that
\[ \frac{\partial(\alpha_i B_j)}{\partial t_{x_k}} = \frac{\partial \alpha_i}{\partial
t_{x_k}}B_j + \alpha_i \frac{\partial B_j}{\partial t_{x_k}} =
\frac{\partial \alpha_i}{\partial t} \frac{\partial t}{\partial
t_{x_k}} B_j + \alpha_i \frac{\partial B_j}{\partial t}
\frac{\partial t}{\partial t_{x_k}} = \alpha_i \alpha_k
\frac{\partial B_j}{\partial t}. \] %
Also, since $\frac{\partial \alpha_i}{\partial x_m} = 0$, we have
\[ \frac{\partial(\alpha_i \alpha_j E_k)}{\partial x_m} = \alpha_i \alpha_j
\frac{\partial E_k}{\partial x_m} .\] %
We can thus carefully calculate the expression $\sum_{i=1}^{3}
\left[ \frac{\partial Re(\mathcal{F}_i)}{\partial t_{x_i}} -
\frac{\partial Im(\mathcal{F}_i)}{\partial x_i} \right]$ in
(\ref{a2_for_elec}) and obtain:
\begin{equation}\label{almost_curlB}
\begin{array}{ll} %
0 = & \left(\alpha_x (\alpha_x - \alpha_y - \alpha_z) - 1 \right)
\left( \frac{\partial E_y}{\partial z} - \frac{\partial
E_z}{\partial y} - \frac{\partial B_x}{\partial t} \right) + \\ %
& \left(\alpha_y (\alpha_x - \alpha_y - \alpha_z) + 1 \right)
\left( \frac{\partial E_z}{\partial x} - \frac{\partial
E_x}{\partial z} - \frac{\partial B_y}{\partial t} \right) + \\ %
& \left(\alpha_z (\alpha_x - \alpha_y - \alpha_z) +1 \right)
\left( \frac{\partial E_x}{\partial y} - \frac{\partial
E_y}{\partial x} - \frac{\partial B_z}{\partial t} \right) .  \\ %
\end{array}
\end{equation}
Denote $ A = \frac{\partial E_y}{\partial z} - \frac{\partial
E_z}{\partial y} - \frac{\partial B_x}{\partial t}, B =
\frac{\partial E_z}{\partial x} - \frac{\partial E_x}{\partial z}
- \frac{\partial B_y}{\partial t}$ and $C = \frac{\partial
E_x}{\partial y} - \frac{\partial E_y}{\partial x} -
\frac{\partial B_z}{\partial t}.$ By taking $\alpha_x= 1$ (and
thus $\alpha_y = \alpha_z = 0$) in (\ref{almost_curlB}), we obtain
$B+C=0$. But this must then be true for any $\alpha_i$'s.
Similarly, the cases $\alpha_y = 1$ and $\alpha_z = 1$ lead
respectively to $-A+C=0$ and $-A+B=0$. From these three equations
we can conclude that $A=B=C=0$, thus proving $\nabla \times
\vec{B} = \frac{\partial \vec{E}}{\partial t}$ as claimed in
(\ref{MaxwellB}).

We now impose the assumption $(a1)$ on the vector
$\vec{\mathcal{F}}$:
\begin{equation}\label{a1_for_elec}
\sum_{i=1}^{3} \left[ \frac{\partial Re(\mathcal{F}_i)}{\partial x_i} -  %
\frac{\partial Im(\mathcal{F}_i)}{\partial
t_{x_i}} \right] = 0 . \end{equation} %
Following the same arguments as before, we obtain
\begin{equation}\label{almost_curlE}
\begin{array}{ll} %
0 = & (\alpha_x+\alpha_z)\left(\frac{\partial B_z}{\partial x} -
\frac{\partial B_x}{\partial z} + \frac{\partial E_y}{\partial t} \right) +  \\ %
& (\alpha_x+\alpha_y)\left(\frac{\partial B_y}{\partial x} -
\frac{\partial B_x}{\partial y}  - \frac{\partial E_z}{\partial t} \right) + \\ %
& (\alpha_y-\alpha_z)\left(\frac{\partial B_z}{\partial y} -
\frac{\partial B_y}{\partial z} - \frac{\partial E_x}{\partial t} \right) . \\ %
\end{array}
\end{equation}
Denote $A = \frac{\partial B_z}{\partial x} - \frac{\partial
B_x}{\partial z} + \frac{\partial E_y}{\partial t}$, $B=
\frac{\partial B_y}{\partial x} - \frac{\partial B_x}{\partial y}
- \frac{\partial E_z}{\partial t}$ and $C= \frac{\partial
B_z}{\partial y} - \frac{\partial B_y}{\partial z} -
\frac{\partial E_x}{\partial t}$. Taking $\alpha_x = 1$ we get
$A+B=0$. Taking $\alpha_y = 1$ we get $B+C=0$. And taking
$\alpha_z = 1$ we get $A-C=0$. These three relations don't suffice
to obtain that $A=B=C=0$. We thus resort to the following "trick".
We rename the original axes by $\tilde{x}: = y, \ \tilde{y}: =z, \
\tilde{z}: =x$. Equation (\ref{almost_curlE}) then becomes
\begin{equation}\label{almost_curlE_2}
\begin{array}{ll} %
0 = & (\tilde{\alpha_x}+\tilde{\alpha_z})\left(\frac{\partial
\tilde{B_z}}{\partial \tilde{x}} - \frac{\partial
\tilde{B_x}}{\partial \tilde{z}} + \frac{\partial
\tilde{E_y}}{\partial t} \right) +  \\ &
(\tilde{\alpha_x}+\tilde{\alpha_y})\left(\frac{\partial
\tilde{B_y}}{\partial \tilde{x}} - \frac{\partial
\tilde{B_x}}{\partial \tilde{y}}  - \frac{\partial
\tilde{E_z}}{\partial t} \right) + \\ &
(\tilde{\alpha_y}-\tilde{\alpha_z})\left(\frac{\partial
\tilde{B_z}}{\partial \tilde{y}} - \frac{\partial
\tilde{B_y}}{\partial \tilde{z}} - \frac{\partial
\tilde{E_x}}{\partial t} \right) . \\
\end{array}
\end{equation}
Now by returning to the original axes we get from
(\ref{almost_curlE_2})
\begin{equation}\label{almost_curlE_3}
\begin{array}{ll} %
0 = & (\alpha_y+\alpha_x)\left(\frac{\partial B_x}{\partial y} -
\frac{\partial B_y}{\partial x} + \frac{\partial E_z}{\partial t} \right) +  \\ %
& (\alpha_y+\alpha_z)\left(\frac{\partial B_z}{\partial y} -
\frac{\partial B_y}{\partial z}  - \frac{\partial E_x}{\partial t} \right) + \\ %
& (\alpha_z-\alpha_x)\left(\frac{\partial B_x}{\partial z} -
\frac{\partial B_z}{\partial x} - \frac{\partial E_y}{\partial t} \right) . \\ %
\end{array}
\end{equation}
Note that the terms in (\ref{almost_curlE_3}) are precisely those
of (\ref{almost_curlE}), which we denoted by $A,B$ and $C$. Taking
for example $\alpha_x = 1$ in (\ref{almost_curlE_3}), we get
$-B+A=0$. Along with the relations obtained before from
(\ref{almost_curlE}), we can now deduce that $A=B=C=0$, thus
proving $\nabla \times \vec{E} = - \frac{\partial
\vec{B}}{\partial t}$ as claimed in (\ref{MaxwellE}). This
completes the proof of the theorem. \sqr

We now turn to mechanics. Suppose that $F$ is any radial force
field, whose source is at the origin of the laboratory frame of
reference $A$. The velocity potential field $\vec{\sigma}$,
created by $F$, is given in our model by
\[
\vec{\sigma} = \left( \begin{array}{l} %
\gamma_s s_x e + \gamma_s \alpha_x i \\ \gamma_s s_y e + \gamma_s
\alpha_y i \\ \gamma_s s_z e + \gamma_s \alpha_z i
\end{array} \right),
\]
where $s_x,s_y,s_z$ denote the components of the velocity (in $A$)
acquired by a particle $m$ in the field $F$,
$s:=(s_x^2+s_y^2+s_z^2)^{\frac{1}{2}}$, and $\gamma_s :=
\frac{1}{\sqrt{1 - s^2}}$ (we take $c\equiv1$). We assume that $F$
is the only force acting, and that $m$ has initial velocity $0$.
In this case, $s_x,s_y,s_z,s$ and $\gamma_s$ are functions only of
$r=(x^2+y^2+z^2)^{\frac{1}{2}}$ and
$t=(t_x^2+t_y^2+t_z^2)^{\frac{1}{2}}$. Thus $\vec{\sigma}$
satisfies assumptions (1) and (2). We take the frame of reference
$B$ to be that of the particle $m$. Thus $\frac{s_{x_i}}{s} =
\frac{v_{x_i}}{v} = \alpha_i.$ \\ In our next theorem, we impose
the assumption $(a2)$ on $\vec{\sigma}$, and obtain a certain
partial differential equation necessarily satisfied by $s$. We
consequently remark that under slow velocity approximations,
gravitational acceleration, which is proportional to
$\frac{1}{r^2},$ satisfies the equation.

\begin{thm}\label{radial}
Suppose that the velocity potential field $\vec{\sigma}$ of any
radial force field $F$ satisfies $(a2)$. Then the velocity $s$ of
a particle in the field $F$ satisfies the equation:
\begin{equation}\label{velocity}
\gamma_s ~ \frac{\partial{s}}{\partial{t}} + s ~
\frac{\partial{\gamma_s}}{\partial{t}} + \frac{\partial
\gamma_s}{\partial r} = 0.
\end{equation}
\end{thm}
\textbf{Proof:} The assumption $(a2)$ on $\vec{\sigma}$ gives
\begin{equation}\label{a2_for_sigma}
\sum_{i=1}^{3} \left( \frac{\partial (\gamma_s s_{x_i})}{\partial
t_{x_i}} + \frac{\partial (\gamma_s \alpha_i)}{\partial x_i}
\right) = 0 .
\end{equation} %
First we calculate the left summand in (\ref{a2_for_sigma}):
\[ \frac{\partial (\gamma_s s_{x_i})}{\partial t_{x_i}} = \gamma_s
\frac{\partial s_{x_i}}{\partial t_{x_i}} + s_{x_i} \frac{\partial
\gamma_s}{\partial t_{x_i}} = \gamma_s \frac{\partial
s_{x_i}}{\partial t} \frac{\partial t}{\partial t_{x_i}} + s_{x_i}
\frac{\partial \gamma_s}{\partial t} \frac{\partial t}{\partial
t_{x_i}}
\]
We previously calculated (see \ref{dtdtxi}) that $\frac{\partial
t}{\partial t_{x_i}} = \frac{t_{x_i}}{t}$. As in the proof of
Theorem \ref{Maxwell}, if we restrict to points in $\M^3$ where
$\frac{t_{x_i}}{t}=\alpha_i$, we obtain
\[ \frac{\partial (\gamma_s s_{x_i})}{\partial t_{x_i}} = \alpha_i
(\gamma_s \frac{\partial s_{x_i}}{\partial t} + s_{x_i}
\frac{\partial \gamma_s}{\partial t}). \] %
Since $\frac{s_{x_i}}{s}$ is constant (and equal to $\alpha_i$),
we have
\[ 0 = \frac{\partial}{\partial t} (\frac{s_{x_i}}{s}) =
\frac{1}{s^2}(\frac{\partial s_{x_i}}{\partial t} s -
\frac{\partial s}{\partial t} s_{x_i}),
\]
and it follows that
\[ \frac{\partial s_{x_i}}{\partial t} = \alpha_i
\frac{\partial s}{\partial t}.
\]
Therefore,
\[ \frac{\partial \!(\! \gamma_s s_{x_i} \!)\!}{\partial t_{x_i}}
\!=\! \alpha_i(\gamma_s \alpha_i \frac{\partial s}{\partial t} +
s_{x_i} \frac{\partial \gamma_s}{\partial t}) \!=\! \alpha_i
(\gamma_s \alpha_i \frac{\partial s}{\partial t} + s \alpha_i
\frac{\partial \gamma_s}{\partial t}) \!=\! \alpha_i^2 (\gamma_s
\frac{\partial s}{\partial t} + s \frac{\partial
\gamma_s}{\partial t}).
\]
Next we calculate the right summand in (\ref{a2_for_sigma}). A
calculation similar to (\ref{dtdtxi}) shows that $\frac{\partial
r}{\partial x_i} = \frac{x_i}{r} .$ Thus
\[ \frac{\partial (\gamma_s \alpha_i)}{\partial x_i} =
\alpha_i \frac{\partial \gamma_s}{\partial x_i} = \alpha_i
\frac{\partial \gamma_s}{\partial r} \frac{\partial r}{\partial
x_i} = \alpha_i \frac{x_i}{r} \frac{\partial \gamma_s}{\partial r}
. \] %
The particle $m$ moves along the radial line where $\frac{x_i}{r}
= \alpha_i$, so that
\[ \frac{\partial (\gamma_s \alpha_i)}{\partial x_i} =
\alpha_i^2 \frac{\partial \gamma_s}{\partial r} . \] %
Returning to equation (\ref{a2_for_sigma}), we obtain:
\[ 0 = \sum_{i=1}^{3} \left( \frac{\partial (\gamma_s s_{x_i})}{\partial
t_{x_i}} + \frac{\partial (\gamma_s \alpha_i)}{\partial x_i}
\right) = \sum_{i=1}^{3} \alpha_i^2 \left( \gamma_s \frac{\partial
s}{\partial t} + s \frac{\partial \gamma_s}{\partial t} +
\frac{\partial \gamma_s}{\partial r} \right) .
\]
But $\sum_{i=1}^{3} \alpha_i^2 = 1$ so we obtain equation
(\ref{velocity}) as claimed. \sqr

\begin{rem}\label{frac1r2}

In non-relativistic mechanics, gravitational force $F$ is given by
$F = ma = \frac{mMG}{r^2}$. Thus the acceleration, $a =
\frac{\partial s}{\partial t}$ is proportional to $\frac{1}{r^2}$.
We will show that subject to non-relativistic approximations,
gravitational acceleration of the form $\frac{1}{r^2}$ satisfies
the equation we obtained in Theorem \ref{radial}. %
So we examine the particular case where $s << c \equiv 1$. Since
\[
\frac{\partial \gamma_s}{\partial t} = \frac{\partial}{\partial t}
\left(\frac{1}{\sqrt{1-s^2}}\right) = - \frac{1}{2}
\frac{1}{(1-s^2)^{\frac{3}{2}}} (-2s) \frac{\partial s}{\partial
t} = s \gamma_s^3 \frac{\partial s}{\partial t} ,
\]
and similarly $\frac{\partial \gamma_s}{\partial r} = s \gamma_s^3
\frac{\partial s}{\partial r},$ we obtain from equation
(\ref{velocity}):
\[ 0 =
\gamma_s \frac{\partial{s}}{\partial{t}} + s
\frac{\partial{\gamma_s}}{\partial{t}} + \frac{\partial
\gamma_s}{\partial r} = \gamma_s \frac{\partial{s}}{\partial{t}} +
s (s \gamma_s^3 \frac{\partial s}{\partial t}) + s \gamma_s^3
\frac{\partial s}{\partial r}.
\]
As $s << 1$, we approximate $\gamma_s \approx 1$ and $s^2 \approx
0$. Subject to these approximations, the above equation gives
\begin{equation}\label{pde_for_s}
\frac{\partial{s}}{\partial{t}} + s \frac{\partial s}{\partial r}
= 0 .
\end{equation}
We want to show that $\frac{\partial{s}}{\partial{t}} = k
\frac{1}{r^2}$, $k \in \Real$, solves (\ref{pde_for_s}). %
Potential energy is given by $E_p = -\frac{m k}{r}$, and kinetic
energy is given by $E_k = \frac{m s^2}{2}$. Thus energy
conservation $E_{k_1} + E_{p_1} = E_{k_2} + E_{p_2}$ gives
$\frac{m s_1^2}{2} + \frac{m k}{r_1} = \frac{m s_2^2}{2} + \frac{m
k}{r_2}$. It follows that $\frac{s_1^2 - s_2^2}{r_2 - r_1} =
\frac{2 k}{r_1 r_2}.$ Thus $ \frac{\partial s^2}{\partial r}(r_0)
:= \lim_{r \rightarrow r_0} \frac{s^2(r) - s^2(r_0)}{r - r_0} =
-\frac{2 k}{r_0^2}.$ Therefore $s \frac{\partial s}{\partial r} =
\frac{1}{2} \frac{\partial s^2}{\partial r} = -\frac{k}{r^2}.$ We
conclude that $\frac{\partial{s}}{\partial{t}} + s \frac{\partial
s}{\partial r} = k \frac{1}{r^2} -\frac{k}{r^2} = 0,$ thus solving
(\ref{pde_for_s}).

\end{rem}

\section{\bf{General remarks}}\label{remarks}

\begin{rem}\label{Fmunu} \textbf{The Tensor Calculus}

We deliberately refrained from using the tensor formalism usually
adapted by physicists. Thus, for example, we wrote explicit matrix
representations of rank 2 tensors, such as $F$, rather than the
more common $F^{\mu \nu}$ notation. Moreover, we didn't develop
the tensor calculus in the $\M$ setting. Our $\mathcal{F}$
undergoing the Lorentz transformation by $\mathcal{F} ' = L \cdot
\mathcal{F} \cdot L$, for example, is justified by the fact that
$\mathcal{F}$ is a contra-variant rank 2 tensor.

\end{rem}

\begin{rem}\label{divergence} \textbf{Assumptions on Anti-symmetric Rank 2 Tensors}

In Theorem \ref{Maxwell} we assumed that the electromagnetic field
vector satisfies $(a1)$ and $(a2)$, and obtained the curl Maxwell
equations. We don't possess $\M$-motivated assumptions such as
$(a1)$ and $(a2)$ at the vector level, yielding the other two
Maxwell equations (in vacuum, with no sources), those of the
divergence:
\[\nabla \cdot \vec{B}= 0 , \hspace{1cm}  and
\hspace{1cm} \nabla \cdot \vec{E}= 0 . \] %
Nevertheless, if we turn to the rank 2 tensor representation of
the electromagnetic field, we do have similar $\M$-motivated
assumption, which we denote $(a3)$ and $(a4)$, that yield both the
divergence equations and the curl equations. We prefer to describe
this as a remark rather than establish it as a theorem, since our
assumptions $(a3)$ and $(a4)$ are not fully justified at this
stage. \\ We write a rank 2 anti-symmetric tensor field as:
\[ \mathcal{T} = \left( \begin{array}{ccc} %
0 & T_{12}e+S_{12}i & T_{13}e+S_{13}i \\ %
-T_{12}e-S_{12}i & 0 & T_{23}e+S_{23}i \\ %
-T_{13}e-S_{13}i & -T_{23}e-S_{23}i & 0
\end{array}\right) . \]
Our new assumptions are
\[ (a3) \hspace{1cm}
\left( \frac{\partial T_{12}}{\partial z} - \frac{\partial
S_{12}}{\partial t_z} \right) - \left( \frac{\partial
T_{13}}{\partial y} - \frac{\partial S_{13}}{\partial t_y} \right)
+ \left( \frac{\partial T_{23}}{\partial x} - \frac{\partial
S_{23}}{\partial t_x} \right) = 0 .
\]
\[ (a4) \hspace{1cm}
\left( \frac{\partial T_{12}}{\partial t_z} - \frac{\partial
S_{12}}{\partial z} \right) - \left( \frac{\partial
T_{13}}{\partial t_y} - \frac{\partial S_{13}}{\partial y} \right)
+ \left( \frac{\partial T_{23}}{\partial t_x} - \frac{\partial
S_{23}}{\partial x} \right) = 0 .
\]
The assumption that the electromagnetic field tensor $\mathcal{F}$
satisfies $(a3)$, translates (after careful calculations and
cancellations) exactly into the divergence Maxwell equation for
$B$:
\[\nabla \cdot \vec{B}= 0 . \]
The assumption that $\mathcal{F}$ satisfies $(a4)$, translates
into
\[ \alpha_x \!\!\left(\!\! \frac{\partial \! E_y}{\partial z} \!-\!
\frac{\partial \! E_z}{\partial y} \!-\! \frac{\partial \!
B_x}{\partial t} \!\!\right)\! + \alpha_y \!\!\left(\!\!
\frac{\partial \! E_z}{\partial x} \!-\! \frac{\partial \!
E_x}{\partial z} \!-\! \frac{\partial \! B_y}{\partial t}
\!\!\right)\! + \alpha_z \!\!\left(\!\! \frac{\partial \!
E_x}{\partial y} \!-\! \frac{\partial \! E_y}{\partial x} \!-\!
\frac{\partial \! B_z}{\partial
t} \!\! \right) \!\!=\! 0 , \] %
from which (by choosing appropriate $\alpha_i$'s, similar to the
proof of Theorem \ref{Maxwell}) we obtain the curl Maxwell
equation for $E$:
\[ \nabla \times \vec{E} = -\frac{\partial
\vec{B}}{\partial t}. \] The electromagnetic field tensor has
another representation (the dual representation), obtained by
interchanging the roles of $E$ and $B$:
\[ \mathcal{F}_1 = \left( \begin{array}{ccc} %
0 & -E_z e + & E_y e + \\ \, %
&[\alpha_x B_y - \alpha_y B_x]i & [\alpha_x B_z - \alpha_z B_x]i
\vspace{0.1cm} \\
E_z e + & 0 & -E_x e + \\ \, %
[\alpha_y B_x - \alpha_x B_y]i & & [\alpha_y B_z - \alpha_z B_y]i
\vspace{0.1cm} \\
-E_y e + & E_x e + &  0 \\ \, %
[\alpha_z B_x - \alpha_x B_z]i & [\alpha_z B_y - \alpha_y B_z]i &
\end{array}\right)  \]
The assumption that $\mathcal{F}_1$ satisfies $(a3)$ translates
into the divergence Maxwell equation for $E$:
\[\nabla \cdot \vec{E}= 0 . \]
And finally, the assumption that $\mathcal{F}_1$ satisfies $(a4)$
translates into
\[ \alpha_x \!\!\left(\!\! \frac{\partial \! B_y}{\partial z} \!-\!
\frac{\partial \! B_z}{\partial y} \!-\! \frac{\partial \!
E_x}{\partial t} \!\!\right)\! + \alpha_y \!\!\left(\!\!
\frac{\partial \! B_z}{\partial x} \!-\! \frac{\partial \!
B_x}{\partial z} \!-\! \frac{\partial \! E_y}{\partial t}
\!\!\right)\! + \alpha_z \!\!\left(\!\! \frac{\partial \!
B_x}{\partial y} \!-\! \frac{\partial \! B_y}{\partial x} \!-\!
\frac{\partial \! E_z}{\partial t}
\!\!\right) \!\!=\! 0 , \] %
from which (by choosing appropriate $\alpha_i$'s) we obtain the
curl Maxwell equation for $B$:
\[ \nabla \times \vec{B} = \frac{\partial
\vec{E}}{\partial t}. \]

\end{rem}

\begin{rem}\label{R6} \textbf{An Alternative Model Based on
$\bf{\mathbb{R}^6}$}

The basic idea of interpreting time in three real dimensions,
could have equally well been mathematically modelled in
$\mathbb{R}^6,$ instead of $\M^3.$ This could be done by replacing
\[ \vec{x} = \left(
\begin{array}{c} xe + ct_xi \\ ye + ct_yi \\ ze + ct_zi
\end{array} \right) \] %
with \[ \vec{x} = \left( \begin{array}{c} x \\ ct_x \\ y \\ ct_y
\\ z \\ ct_z \\
\end{array} \right) , \] %
and likewise other vectors; Matrices in $\M_{3 \times 3}$ should
then translate to matrices in $ \mathbb{R}_{6 \times 6}$ by
replacing each $\alpha e + \beta i$ entry with a $2\times 2$ block
of the form
\[\left( \begin{array}{cc}
\alpha & \beta \\
\beta & \alpha
\end{array} \right).
\]
For example, our Lorentz transformation matrix $L$ would be
replaced by
\[ \left( \!\!\! \begin{array}{llllll}
1  \!\!+\!\! (\!\gamma \!\!-\!\! 1 \!)\alpha_x^2 \!&\!
-\beta\gamma\alpha_x^2 \!&\! \alpha_x\alpha_y(\!\gamma \!\!-\!\! 1
\!) \!&\! -\alpha_x\alpha_y\beta\gamma \!&\!
\alpha_x\alpha_z(\!\gamma \!\!-\!\! 1 \!) \!&\!
-\alpha_x\alpha_z\beta\gamma \\
-\beta\gamma\alpha_x^2 \!&\! 1 \!\!+\!\! (\!\gamma \!\!-\!\! 1 \!)
\alpha_x^2 \!&\! -\alpha_x\alpha_y\beta\gamma \!&\!
\alpha_x\alpha_y(\! \gamma \!\!-\!\! 1 \!) \!&\!
-\alpha_x\alpha_z\beta\gamma \!&\! \alpha_x \alpha_z(\!\gamma
\!\!-\!\! 1 \!) \\
\alpha_x\alpha_y(\!\gamma \!\!-\!\!1\!) \!&\!
-\alpha_x\alpha_y\beta\gamma \!&\! 1 \!\!+\!\! (\!\gamma \!\!-\!\!
1 \!)\alpha_y^2 \!&\! -\beta\gamma\alpha_y^2 \!&\!
\alpha_y\alpha_z(\!\gamma \!\!-\!\! 1 \!) \!&\!
-\alpha_y\alpha_z\beta\gamma \\
-\alpha_x\alpha_y\beta\gamma \!&\! \alpha_x\alpha_y( \! \gamma
\!\!-\!\! 1 \! ) \!&\! -\beta\gamma\alpha_y^2 \!&\! 1 \!\!+\!\! (
\! \gamma \!\!-\!\! 1 \! )\alpha_y^2 \!&\!
-\alpha_y\alpha_z\beta\gamma
\!&\! \alpha_y\alpha_z( \! \gamma \!\!-\!\! 1 \! ) \\
\alpha_x\alpha_z(\! \gamma \!\!-\!\! 1 \!) \!&\!
-\alpha_x\alpha_z\beta\gamma \!&\! \alpha_y\alpha_z(\! \gamma
\!\!-\!\! 1 \!) \!&\! -\alpha_y\alpha_z\beta\gamma \!&\! 1
\!\!+\!\! (\!\gamma \!\!-\!\! 1 \!)\alpha_z^2 \!&\!
-\beta\gamma\alpha_z^2
\\
-\alpha_x\alpha_z\beta\gamma \!&\! \alpha_x\alpha_z( \! \gamma
\!\!-\!\! 1 \! ) \!&\! -\alpha_y\alpha_z\beta\gamma \!&\!
\alpha_y\alpha_z( \! \gamma \!\!-\!\! 1 \! ) \!&\!
-\beta\gamma\alpha_z^2 \!&\! 1 \!\!+\!\! ( \! \gamma \!\!-\!\! 1
\! ) \alpha_z^2
\end{array} \!\!\! \right) \!. \]

Our motivation for using $\M^3$, is that the new mathematics seems
to inspire the physics. For example, assumptions (a1) and (a2)
above, which arise from a pure mathematical point of view, lead to
the physical results in Theorems \ref{Maxwell} and \ref{radial}.
Nevertheless, the $\M^3$ model has it's disadvantages:
\begin{enumerate}

\item
There is no Lorentz group (see Remark \ref{Lorentz} below).

\item
We can't express rotations.

\end{enumerate}

It is possible that the $\mathbb{R}^6$ model could overcome these
problems. We thus don't rule out the possibility of converting to
$\mathbb{R}^6$, or reestablishing our theory based on the
interplay between the $\mathbb{R}^6$ and $\M^3$ models, and
possibly other "pseudo-complex" mathematical models (see Remark
\ref{otherM} below).

\end{rem}

\begin{rem}\label{Lorentz} \textbf{Absence of the Lorentz Group}

In our $\M^3$ model, Lorentz transformation matrices cannot be
composed. The reason for this is as follows. Suppose that a vector
$\vec{x}_{AB}$ represents a physical event in a certain frame of
reference $A$, with respect to a second frame of reference $B$.
Let $L_{AB}$ denote the appropriate Lorentz transformation matrix.
Now suppose that $C$ is a third frame of reference. In general, it
is meaningless to have $L_{BC}$ act on $L_{AB} (\vec{x}_{AB})$,
since $L_{AB} (\vec{x}_{AB})$ is not equal to $\vec{x}_{BC}$
(recall that $\vec{x}_{BC}$ contains the $\alpha$'s of the
relative velocity between $B$ and $C$). In particular, it is not
true that $L_{BC} \circ L_{AB} = L_{AC}$. Thus, our model doesn't
give rise to an analog of the Lorentz group.

\end{rem}

\begin{rem}\label{otherM} \textbf{Other Possible Pseudo-Complex
Models}

We restrict our attention to a single space dimension, denoted by
the $x$ axis, in order to present the following observation. The
regular Lorentz transformation is given by
\[
\left( \begin{array}{cc} %
\gamma & -\beta\gamma \\ %
-\beta \gamma & \gamma
\end{array} \right) %
\left( \begin{array}{c} x \\ ct \end{array} \right) = %
\left( \begin{array}{c} \gamma x - \beta\gamma ct \\ -\beta\gamma
x + \gamma ct \end{array} \right) .
\]
In our model, we coupled the $x$ and $t$ coordinates into a single
"pseudo-complex" ($\M$) coordinate, $xe+cti$. We then replaced the
regular Lorentz transformation above with:
\[ (\gamma e + \beta\gamma i)(xe+cti) =
(\gamma x - \beta\gamma ct)e + (-\beta\gamma x + \gamma ct)i.
\]
But in doing so, we made a (seemingly) arbitrary choice of signs
in the following expression:
\[ (\pm \gamma e \pm \beta\gamma i)(xe+cti) =
(\gamma x - \beta\gamma ct)e \pm (-\beta\gamma x + \gamma ct)i .
\]
Any other combination of $\pm$ signs above is possible, leading to
a different definition of multiplication, and thus a totaly
different "pseudo-complex" model. We demonstrate some examples in
a multiplication chart:
\[
\begin{array}{ccccccccc}
\!\!\! \left. \begin{array}{c} \\ ee \\ ei \\ ie \\ ii \end{array}
\right|
\! \left. \begin{array}{c} \!+\!+\!+\! \\ e \\ i \\ -i \\ -e
\end{array} \right|
\! \left. \begin{array}{c} \!+\!-\!+\! \\ e \\ i \\ i \\ e
\end{array} \right|
\! \left. \begin{array}{c} \!-\!+\!+\! \\ -e \\ -i \\ -i \\ -e
\end{array} \right|
\! \left. \begin{array}{c} \!-\!-\!+\! \\ -e \\ -i \\ i \\ e
\end{array} \right|
\! \left. \begin{array}{c} \!+\!+\!-\! \\ e \\ -i \\ i \\ -e
\end{array} \right|
\! \left. \begin{array}{c} \!+\!-\!-\! \\ e \\ -i \\ -i \\ e
\end{array} \right|
\! \left. \begin{array}{c} \!-\!+\!-\! \\ -e \\ i \\ i \\ -e
\end{array} \right|
\! \left. \begin{array}{c} \!-\!-\!-\! \\ -e \\ i \\ -i \\ e
\end{array} \right|
\end{array}\]
The $+++$ choice is our $\M$, while the $+-+$ option for example
is a commutative associative algebra with a unit! Note though that
it is impossible to recover the regular complex numbers
$\mathbb{C}$.

We should point out the fact that all these possible models seem
to lead back to the same $\mathbb{R}^6$ model, discussed in Remark
\ref{R6}. It is worth investigating mathematically these different
possible models, and seeing what they give rise to on the physical
side.
\end{rem}

\textbf{Acknowledgements:} We wish to thank Eran Rosenthal and
Larry Horwitz for helpful discussions. This research was supported
by the Graduate School of the Technion - Israel Institute of
Technology.

\end{document}